\documentclass{article}
\usepackage{spconf,amsmath,graphicx,hyperref}

\usepackage{xcolor}
\usepackage[table]{xcolor}
\usepackage{listings}
\usepackage{booktabs}
\usepackage{multirow, multicol}
\usepackage[skins,breakable]{tcolorbox}

\usepackage{tikz}
\usetikzlibrary{tikzmark} 

\definecolor{toolgray}{HTML}{999999}
\definecolor{geminiblue}{HTML}{87CEFA}
\definecolor{humansred}{HTML}{F18080}

\definecolor{riceyellow}{HTML}{FFFFFB}
\definecolor{boxborder}{HTML}{EEE9DA}

\usepackage[dvipsnames]{xcolor}

\lstdefinestyle{jsonoutput}{
    basicstyle=\ttfamily\small,
    columns=fullflexible,
    keepspaces=true,
    showstringspaces=false,
    breaklines=true,
    frame=none,
    escapeinside={(*@}{@*)},
}

\usepackage{cite}
\usepackage{amsmath,amssymb,amsfonts}
\usepackage{algorithmic}
\usepackage{graphicx}
\usepackage{textcomp}
\usepackage{xcolor}
\def\BibTeX{{\rm B\kern-.05em{\sc i\kern-.025em b}\kern-.08em
    T\kern-.1667em\lower.7ex\hbox{E}\kern-.125emX}}

\title{Conversational Voice Aesthetic Model with \\Reinforcement Learning from Human Listeners}

\twoauthors
 {Xilin Jiang\sthanks{Work done during an internship in Meta.}}{Columbia University\\New York, NY, USA}
 {Shun Zhang, Tejas Jayashankar, Yinghao Aaron Li, Osama Hanna}{Meta Superintelligence Labs\\Menlo Park, CA, USA}

\begin{document}
\ninept
\maketitle
\begin{abstract}
We introduce Conversational Voice Aesthetic Model, a speech large language model for describing the voice aesthetics of real or synthetic speech responses in natural conversational contexts. Given a context and a response speech, CVAM describes salient moments that characterize the voice and predicts nine categorical attributes spanning gender, pitch, pacing, emotion, and delivery. The key challenge lies in perceptual fields such as emotion and delivery, which are inherently subjective and lack definitive ground truth. Therefore, we collect $\sim$10 human annotations for each of 3k real and synthetic responses derived from the CANDOR corpus. CVAM is supervised finetuned on synthesized aesthetic descriptions and labels, then optimized with Group Relative Policy Optimization on human judgments. Experiments show that CVAM better agrees with human listeners than Gemini 3.1 Pro and open-source speech LLMs, and outperforms single-human-vs.-rest agreement. Together, we demonstrate the importance of grounding voice aesthetics in human perception and propose a principled framework for human alignment.
\end{abstract}

\section{Introduction}
\textbf{\textit{Voice}} is a fundamental medium of human communication \cite{fitch2000evolution, stivers2009universals}. Long before writing systems emerged, people conveyed meaning, intent, and feeling through voice, and spoken interaction remains a natural way humans connect with one another and, increasingly, with machines. As conversational agents mature, the voice itself, not merely the words, becomes the \textbf{\textit{aesthetic}} object to describe: from physical aspects such as \textit{pitch} and \textit{pacing} to contextual aspects such as \textit{emotion} and \textit{delivery}. Human-faithful voice description can facilitate both conversation understanding and expressive voice synthesis.

\begin{figure}[!ht]
    \centering
    \includegraphics[width=\linewidth]{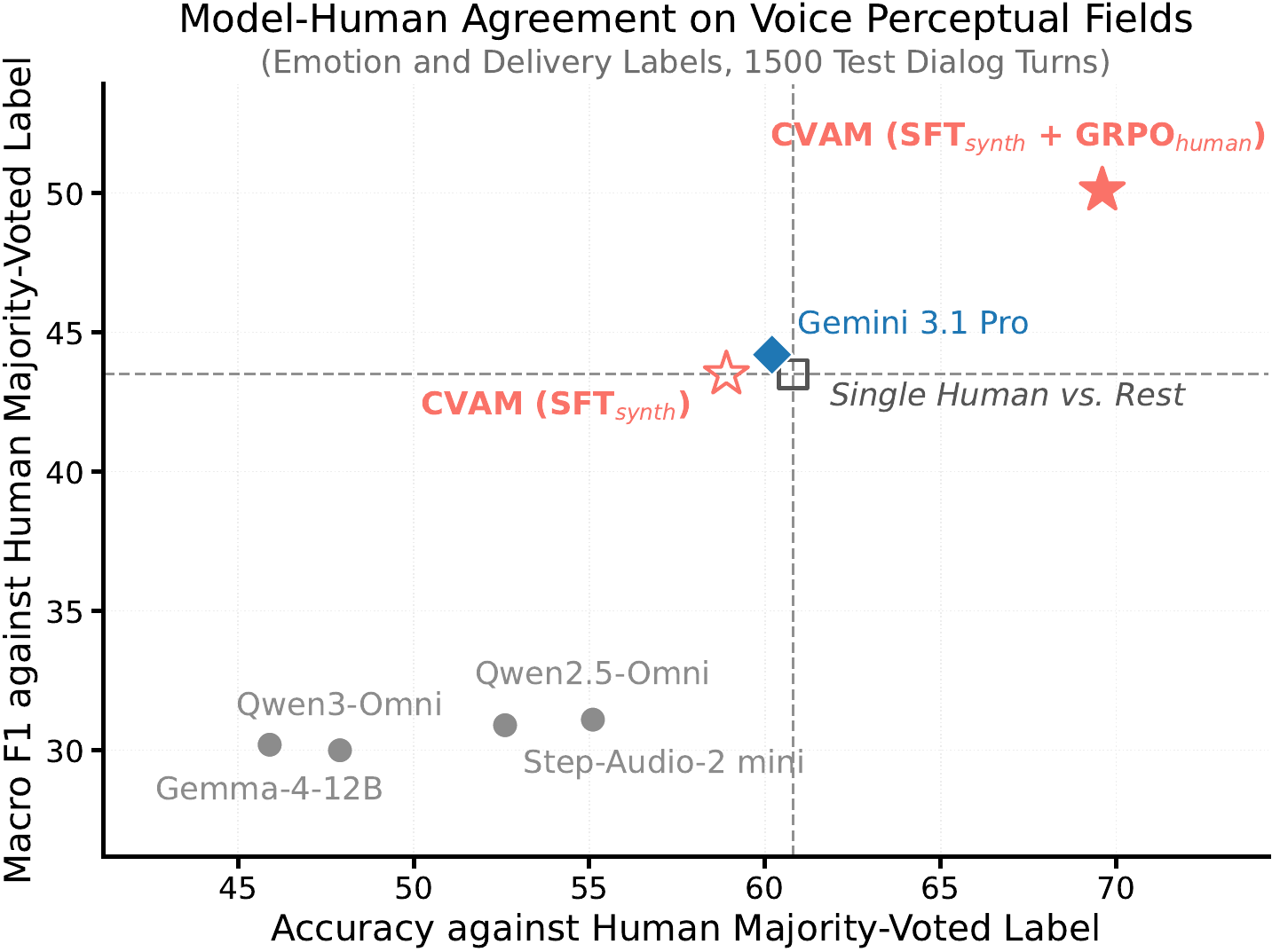}
    \caption{With supervised finetuning on synthetic annotation and reinforcement learning on human crowd judgments, CVAM better agrees with the human crowd on emotion and delivery labels than Gemini 3.1 Pro and evaluated open-source speech LLMs in both accuracy and macro F1 against human majority-voted labels. Notably, CVAM also exceeds the single-human-vs-rest agreement.}
    \label{fig:takeaway}
    \vspace{-0.6cm}
\end{figure}

Voice measures, descriptors, and judges have evolved from signal methods to neural predictors and, more recently, to speech large language models. Early measures such as PESQ \cite{rix2001pesq} and POLQA \cite{beerends2013polqa} assess acoustic signals with perceptually motivated signal processing. Neural models such as MOSNet \cite{lo2019mosnet}, DNSMOS \cite{reddy2021dnsmos}, and UTMOSv2 \cite{baba2024utmosv2} are trained to predict speech quality, while Audiobox-Aesthetics \cite{tjandra2025aes} extends prediction to multiple aesthetic dimensions. Speech LLMs enable more accurate and interpretable voice description. AudioJudge \cite{manakul2025audiojudge}, WavReward \cite{ji2025wavreward}, SpeechJudge \cite{zhang2025speechjudge}, and GSRM \cite{shen2026gsrm} assess speech attributes and, in some cases, generate explicit reasoning. Yet, they largely judge speech utterances without conversational context, even though attributes such as emotion and delivery depend on preceding turns. Moreover, with advances in text-to-speech
\cite{li2023styletts2,anastassiou2024seed,qwen3tts},
speech-to-speech
\cite{defossez2024moshi,openai2024gpt4o,zeng2024glm4},
and voice agents
\cite{roy2026personaplex,dindar2026sympatheiaemotionallyadaptivevoice,jiang2026cocktail},
a voice aesthetic model is increasingly expected to describe both real and synthetic speech with the same, fair standard.

Another, yet fundamental, challenge is human alignment. Perceptual attributes like emotion, warmth, and confidence are subjective and lack definitive ground truth. Individual listeners may be biased and inconsistent, while multiple listeners can form different yet valid judgments when taken collectively. Voice aesthetic modeling therefore requires human crowd supervision and objectives that capture both majority judgments and the distribution of listener votes.

These gaps motivate our development of \textbf{Conversational Voice Aesthetic Model (CVAM)}, grounded in human perception. Our human listening study of 3k responses from the CANDOR corpus \cite{reece2023candor} first reveals substantial misalignment between existing speech LLMs and human voice-aesthetics judgments. To address this discrepancy, we introduce a two-stage post-training framework: supervised finetuning (SFT) on 30k samples with synthesized descriptions and labels, followed by reinforcement learning (RL) on 1.5k human-annotated samples with a distribution-based reward over listener votes. The resulting CVAM predicts structured voice aesthetic attributes in conversational context, outperforming all evaluated speech LLMs in human alignment and surpassing single-human-vs.-rest agreement. Together, we establish a principled framework for human-aligned conversational voice aesthetic modeling.

\section{Method}
CVAM is based on Qwen2.5-Omni \cite{xu2025qwen25omni}, a 7B speech LLM with strong speech understanding abilities. Fig.~\ref{fig:model} illustrates its input and output. This work considers four voice aesthetic fields: \textbf{\textit{pitch}}, \textbf{\textit{pacing}}, \textbf{\textit{emotion}}, \textbf{\textit{delivery}}, together with the speaker's \textbf{\textit{gender}}. Pitch and pacing are objective fields measured by speech analysis tools, while emotion and delivery are perceptual fields judged by humans. Each aesthetic field comprises an evidence-grounded description and two categorical labels, yielding nine labels including gender.

\subsection{Data Preparation} \label{sec:data}

\noindent
\textbf{Conversation Preprocessing.}
We derive conversational turns from CANDOR \cite{reece2023candor}, a naturalistic English conversation corpus. Each sample contains the preceding speaker's context audio and the target speaker's response audio, both restricted to 3--30 seconds. We remove backchannels and apply a semantic filter (Qwen3-32B \cite{yang2025qwen3}) to exclude politics, race, religion, personal information, survey logistics, payment, and other sensitive topics. This yields 10k+0.5k training and 0.5k testing turns, with no speaker or content overlap. Qwen3-TTS \cite{qwen3tts} supplements each real response with two synthetic versions of the same transcript (effectively tripling the dataset): a voice clone (enrollment sampled from a different turn of the same speaker) and a voice design, whose voice prompt is written by Qwen3-Omni \cite{xu2025qwen3omni} given the context audio, uniformly \& independently sampled emotion, arousal, warmth, and confidence labels, and the speaker's gender and age. Voice clone is mostly neutral, but voice design improves the coverage of rare emotion and delivery.

\begin{figure}[!th]
    \centering
    \includegraphics[width=\linewidth]{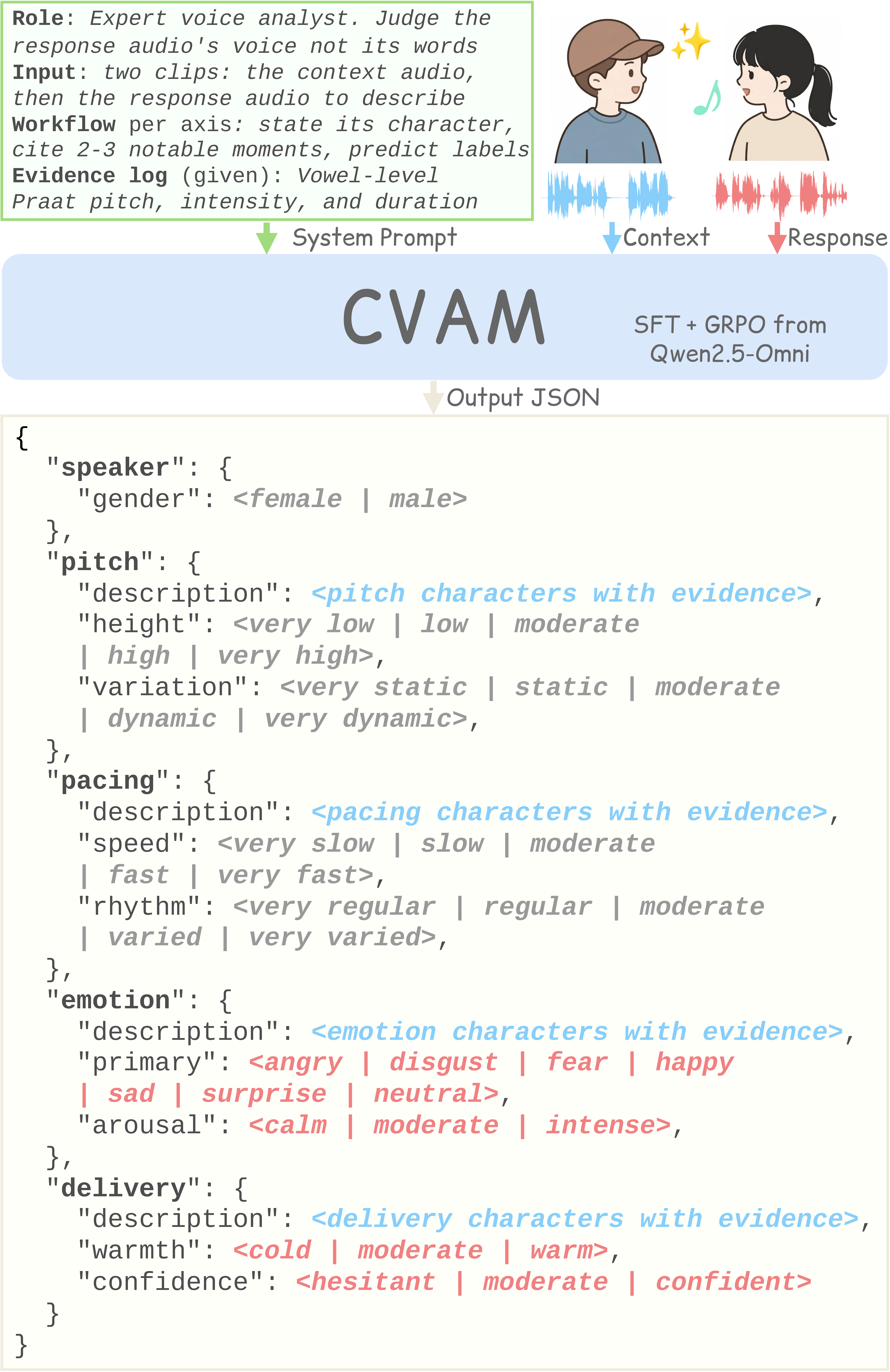}
    \caption{CVAM receives context \& response speech and outputs JSON descriptions and labels for the response's gender, pitch, pacing, emotion, and delivery. Colors indicate target sources:
    \textcolor{toolgray}{\textit{self-reported gender and tool-measured objective labels}},
    \textcolor{geminiblue}{\textit{LLM-written descriptions}}, and
    \textcolor{humansred}{\textit{perceptual labels from an LLM for SFT or humans for GRPO}}.}
    \label{fig:model}
\end{figure}

\noindent\rule{\linewidth}{0.4pt}

\noindent
\textbf{Tool Measures \& Objective Labels.}
We extract pitch (F0) with Praat \cite{boersma2001praat}, transcribe with faster-whisper-large-v3 \cite{radford2023whisper}, and obtain phoneme-level alignment with NeMo Forced Aligner \cite{kuchaiev2019nemo} and CMUdict. Label cutoffs are: pitch height (mean F0, Hz) at $[100,150,200,250]$ (\textit{very low}--\textit{very high}); pitch variation (F0 std, semitones) at $[2,3,4,5]$ (\textit{very static}--\textit{very dynamic}); pacing speed (phonemes per second over the speech span) at $[7,9,11,13]$ (\textit{very slow}--\textit{very fast}); and pacing rhythm (local phoneme rate std, 1\,s window, 0.25\,s hop) at $[2,3,4,5]$ (\textit{very regular}--\textit{very varied}). Each sample also yields an \textit{evidence log} of vowel-level measures. CVAM receives the tool measures and evidence log by default, making the prediction of objective fields largely trivial. We investigate whether the tool measures also improve the prediction of perceptual fields.

\noindent\rule{\linewidth}{0.4pt}

\noindent
\textbf{Synthetic Descriptions \& Labels.}
Human annotation, on the scale required to fully train the model, is difficult to obtain. We therefore first annotate 30k samples (10k turns$\times$3 versions) with Gemini 3.1 Pro \cite{google2026gemini31pro} (for SFT), using four specialized calls per sample, and leave humans to annotate 1.5k samples (labels only, for RL). Pitch and pacing calls receive the response audio and tool measures to produce descriptions that \textit{explain} the objective labels. Emotion and delivery calls receive context and response audio, without potentially biasing tool measures, and jointly produce descriptions and perceptual labels. Outputs are merged into the JSON format in Fig.~\ref{fig:model}.

\begin{figure}[!th]
    \centering
    \includegraphics[width=\columnwidth]{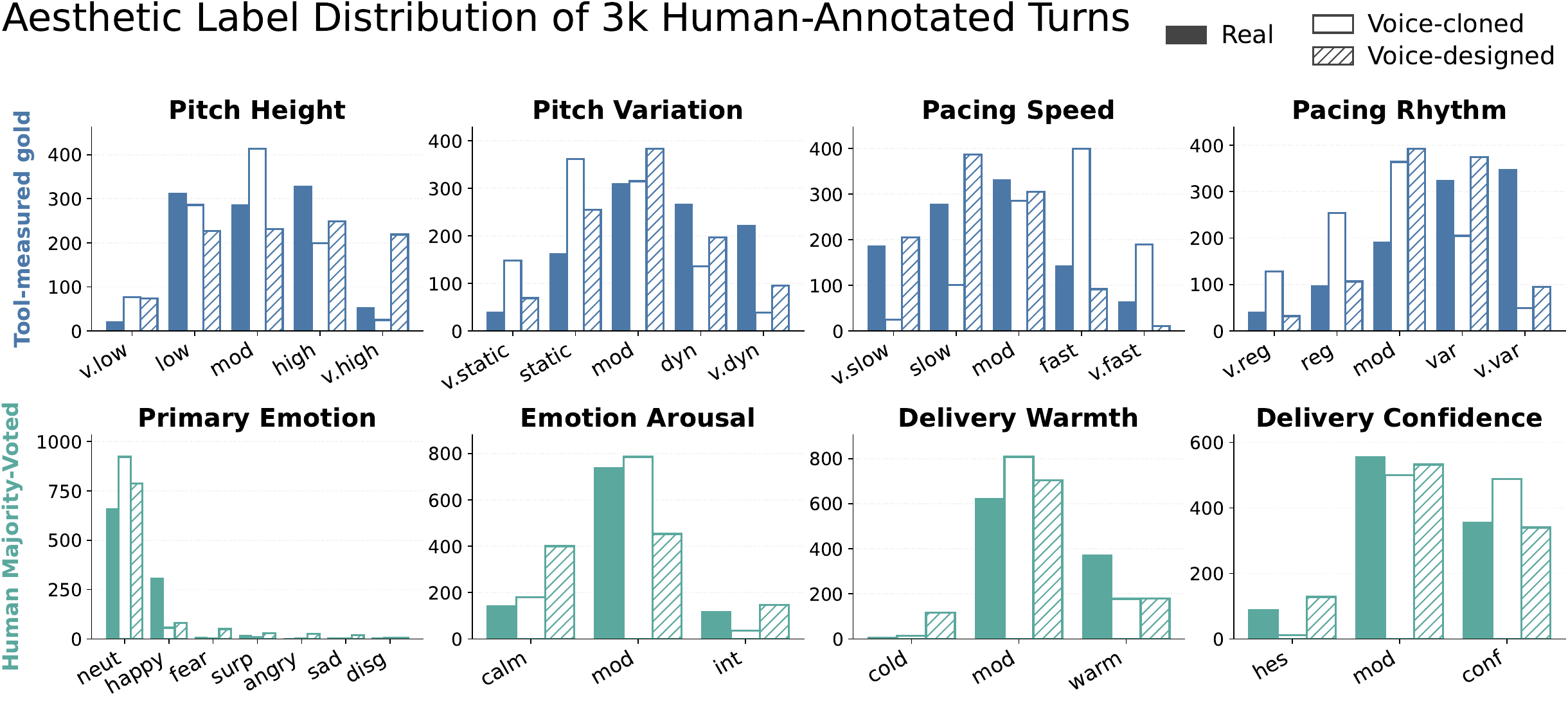}
    \caption{Distributions of the objective labels (measured by speech analysis tools) and perceptual labels (human majority-voted) of 3k human-annotated speech responses from the CANDOR corpus.}
    \label{fig:distribution}
    \vspace{-0.7cm}
\end{figure}

\noindent\rule{\linewidth}{0.4pt}

\noindent
\textbf{Human-Annotated Perceptual Labels.}
We collect human judgments on 1.5k testing and 1.5k training turns. Listeners compare two response versions (real vs.\ cloned, real vs.\ designed, cloned vs.\ designed) sharing identical context and annotate each version separately, with 5 raters per pair. Since each version appears twice across three pairs, this accumulates to $N\approx10$ votes after invalid entries are discarded. Testing and training splits respectively average 9.78 and 9.71 votes per response (both medians: 10; 79.7\% and 73.1\% with at least ten), totaling 14{,}664 and 14{,}566 rater--response judgments per field. Perceptual labels are defined as the human majority-voted. Fig.~\ref{fig:distribution} shows their distributions by real/synthetic response versions.

\subsection{Two-Stage Training}

Training only on human data is often data-starved, whereas training only on synthetic data inherits the LLM annotator's miscalibration. We therefore first supervised finetune (SFT) CVAM on 30k synthetic samples (Sec.~\ref{sec:data}), targeting the full JSON of four descriptions and nine labels to establish the output format and aesthetic description. The second stage aligns predicted labels to human votes with reinforcement learning (RL). RL allows learning multiple votes rather than the mode: SFT on a point human label discards vote variation, whereas distribution-based RL rewards can align sampled predictions with the full vote distribution (also without the need for human-written descriptions). Results (Sec.~\ref{sec:results}) show that two-stage training outperforms human- or synthetic-only training, or SFT on both.

Both stages tune LoRA \cite{hu2022lora} adapters (rank 32, $\alpha=64$) on all thinker linear layers, freezing the audio encoder and aligner and disabling the talker head. We train for one epoch with a cosine schedule and 5\% warmup in bfloat16. SFT and RL respectively use learning rates of $10^{-4}$ and $10^{-5}$ on $2\times8$ and $4\times8$ A100s. All experiments use \texttt{ms-swift} \cite{zhao2025swift}, with vLLM \cite{kwon2023vllm} accelerator.

\noindent
\textbf{GRPO} \cite{shao2024deepseekmath} scores the nine predicted labels parsed from the generated JSON. For each input ($x$), we sample $G=16$ rollouts ($y_i$) from the rollout policy ($\pi_{\mathrm{old}}$) at temperature 1.0. We maximize $\mathcal{J}(\theta)=$
\begin{equation}
\begin{aligned}
\mathbb{E}\Big[
\tfrac{1}{G}\sum_{i=1}^{G}
\min\big(\rho_i A_i, \mathrm{clip}(\rho_i,1\pm\epsilon)A_i\big)
\Big]
-\beta\,\mathbb{D}_{\mathrm{KL}}
[\pi_\theta\|\pi_{\mathrm{ref}}]
\end{aligned}
\label{eq:grpo}
\end{equation}
where the expectation covers training inputs and sampled rollouts,
$\rho_i$ denotes the importance ratio between the trainable
($\pi_\theta$) and rollout ($\pi_{\mathrm{old}}$) policies,
and $\epsilon$ is the clipping threshold.
The advantage $A_i=(R_i-\bar R)/(\sigma_R+\delta)$ uses
rollout reward ($R_i$), group mean ($\bar R$), standard deviation
($\sigma_R$), and numerical stabilizer ($\delta$).
Training is fully on-policy, with KL divergence
($\mathbb{D}_{\mathrm{KL}}$) against the frozen SFT policy
($\pi_{\mathrm{ref}}$) weighted by $\beta=0.04$.
Mean subtraction makes reward-driven updates depend on
within-group differences.

Gender and the four objective labels use accuracy reward on deterministic point labels, contributing $5/9$ of the total reward. The four perceptual fields contribute the other $4/9$. Nevertheless, since predictions of gender and objective labels are already nearly perfect, their rewards vary little across rollouts and contribute little to the relative advantage. Thus, nearly all learning signals come from perceptual labels. We contrast two perceptual reward families: Accuracy-based rewards compare against a random single human's label, the majority-voted label, or the fraction of humans sharing the predicted label. Distribution-based rewards match group label frequencies to human vote distributions ($p$), separately for each perceptual field:
\begin{equation}
\begin{aligned}
r_i = \phi\!\left(p(c_i),\tilde q_{-i}(c_i)\right), \text{ where }
\tilde q_{-i}(c_i)
= \frac{n(c_i)-1+\lambda}{G-1+\lambda K}
\end{aligned}
\label{eq:dist}
\end{equation}
where $r_i$ the perceptual reward, $c_i$ the predicted label of $i$-th rollout,
$n(c_i)$ its group count, $G=16$ the group size, and $K$ the number of distinct human
or predicted labels. The estimate ($\tilde q_{-i}$) excludes $i$-th rollout itself and uses
Laplace smoothing ($\lambda=1$). The intuition is to align predictions with the diversity of human judgments rather than simply reinforce the majority label. Each rollout is rewarded by comparing human support for its predicted label with that label's frequency among the other rollouts, favoring underproduced labels and discouraging overproduced ones. Excluding the rollout itself prevents its prediction from inflating the estimated frequency, while smoothing handles labels absent from the remaining rollouts.

Various distribution distances ($\mathcal{D}$) can be adopted as rewards that align the rollout distribution ($q$) with the human vote distribution ($p$). Since each rollout predicts a single label $c$, we derive the reward ($\phi$) as $-\partial \mathcal{D}(p,q)/\partial q(c)$, so that each policy gradient update locally reduces $\mathcal{D}$. These include TVD-inspired $\phi(p,q)=\mathrm{sign}(p-q)$; forward-KL-inspired $\phi(p,q)=\min(p/q,\kappa)$; and reverse-KL-inspired $\phi(p,q)=\mathrm{clip}(\log(\max(p,\eta)/q),-\kappa,\kappa)$, with reward cap ($\kappa=5$) and probability floor ($\eta=0.01$). Sec.~\ref{sec:results} examines different rewards' accuracy--coverage (macro F1) trade-off.

\begin{table}[h]
\centering
\small
\setlength{\tabcolsep}{3pt}
\resizebox{\columnwidth}{!}{%
\begin{tabular}{l | cc cc cc cc | cc}
\toprule
\multirow{2}{*}{System} & \multicolumn{2}{c}{Emotion} & \multicolumn{2}{c}{Arousal} & \multicolumn{2}{c}{Warmth} & \multicolumn{2}{c}{Confidence} & \multicolumn{2}{c}{Avg. Perceptual} \\
\cmidrule(lr){2-3} \cmidrule(lr){4-5} \cmidrule(lr){6-7} \cmidrule(lr){8-9} \cmidrule(lr){10-11}
 & Acc & F1 & Acc & F1 & Acc & F1 & Acc & F1 & Acc & F1 \\
\midrule
\multicolumn{11}{c}{Baseline Speech Large Language Models} \\ \midrule
Qwen2.5-Omni \cite{xu2025qwen25omni} & 70.7 & 27.6 & 64.1 & 38.1 & 42.1 & 28.5 & 43.5 & 30.3 & 55.1 & 31.1 \\
Qwen3-Omni \cite{xu2025qwen3omni} & 75.1 & 29.8 & 41.3 & 39.1 & 37.1 & 24.7 & 38.3 & 26.3 & 47.9 & 30.0 \\
Step-Audio-2 mini \cite{wu2025stepaudio2} & 51.7 & 24.1 & 61.3 & 34.7 & 43.9 & 29.7 & 53.4 & 35.2 & 52.6 & 30.9 \\
Gemma-4-12B \cite{gemmateam2026gemma4} & 55.5 & 25.2 & 39.3 & 29.7 & 44.0 & 29.7 & 44.7 & 36.2 & 45.9 & 30.2 \\
Gemini 3.1 Pro \cite{google2026gemini31pro} & 74.1 & 37.9 & 55.9 & 46.3 & 59.8 & 44.4 & 51.0 & 48.4 & 60.2 & 44.2 \\
$^\dagger$Gemini 3.1 Pro & 73.9 & 35.8 & 54.5 & 45.3 & 66.5 & 51.6 & 45.9 & 44.0 & 60.2 & 44.2 \\
\midrule
\multicolumn{11}{c}{Conversational Voice Aesthetic Model (CVAM)} \\ \midrule
SFT$_{\textit{synth}}$ & 73.3 & 36.9 & 54.3 & 45.4 & 60.0 & 46.8 & 46.9 & 44.2 & 58.6 & 43.3 \\
$^\dagger$SFT$_{\textit{synth}}$ & 72.9 & 36.1 & 55.0 & 46.0 & 60.1 & 47.1 & 47.5 & 44.8 & 58.9 & 43.5 \\
$^\dagger$SFT$_{\textit{human}}$ & 72.1 & 25.5 & 71.8 & 42.0 & 66.9 & 43.0 & 62.4 & 42.0 & 68.3 & 38.1 \\
$^\dagger$GRPO$_{\textit{human}}^{\textit{KL}}$ & 64.4 & 26.7 & 65.7 & 42.5 & 59.0 & 40.1 & 43.1 & 29.6 & 58.1 & 34.7 \\
$^\dagger$SFT$_{\textit{synth}}$ + SFT$_{\textit{human}}$ & 83.9 & 33.3 & 69.7 & 50.7 & 71.9 & 47.0 & 55.5 & 48.5 & 70.2 & 44.9 \\
$^\dagger$\textbf{SFT$_{\textit{synth}}$ + GRPO$_{\textit{human}}^{\textit{Frac}}$} & \textbf{88.1} & 33.0 & \textbf{78.8} & 52.5 & \textbf{78.7} & \textbf{53.6} & 65.7 & \textbf{51.1} & \textbf{77.8} & 47.6 \\
$^\dagger$\textbf{SFT$_{\textit{synth}}$ + GRPO$_{\textit{human}}^{\textit{KL}}$} & 83.9 & \textbf{40.5} & 75.3 & \textbf{59.3} & 66.0 & 51.9 & 53.1 & 48.6 & 69.6 & \textbf{50.1} \\
\midrule
\textit{Single Human vs. Rest} & \textit{70.7} & \textit{32.5} & \textit{59.7} & \textit{50.3} & \textit{58.5} & \textit{45.0} & \textit{54.2} & \textit{46.2} & \textit{60.8} & \textit{43.5} \\
\bottomrule
\end{tabular}}
\caption{Agreement with human listeners on the perceptual fields. Accuracy and macro-F1 are against the majority-voted ($N\approx10$) label for the perceptual fields. $^\dagger$\textit{Tools measure} provides the pitch and pacing labels and the evidence log. \textit{Single Human} scores \textit{each} one of the $N$ raters against the other $N-1$ and reports the average.}
\label{tab:perceptual}
\end{table}

\begin{table}[h]
\centering
\small
\setlength{\tabcolsep}{3pt}
\resizebox{\columnwidth}{!}{%
\begin{tabular}{l | cc cc cc cc cc}
\toprule
\multirow{2}{*}{System} & \multicolumn{2}{c}{Gender} & \multicolumn{2}{c}{PitchHgt} & \multicolumn{2}{c}{PitchVar} & \multicolumn{2}{c}{Speed} & \multicolumn{2}{c}{Rhythm} \\
\cmidrule(lr){2-3} \cmidrule(lr){4-5} \cmidrule(lr){6-7} \cmidrule(lr){8-9} \cmidrule(lr){10-11}
 & Acc & F1 & Acc & F1 & Acc & F1 & Acc & F1 & Acc & F1 \\
\midrule
\multicolumn{11}{c}{Baseline Speech Large Language Models} \\ \midrule
Qwen2.5-Omni & 86.0 & 84.2 & 30.7 & 9.5 & 32.6 & 12.5 & 30.9 & 9.4 & 15.9 & 8.2 \\
Qwen3-Omni & 94.4 & 93.9 & 41.1 & 26.2 & 29.4 & 20.1 & 32.8 & 18.1 & 15.8 & 9.4 \\
Step-Audio-2 mini & 76.2 & 75.4 & 30.7 & 9.8 & 24.9 & 13.8 & 30.9 & 10.1 & 23.7 & 13.6 \\
Gemma-4-12B & 72.1 & 66.2 & 32.7 & 16.9 & 26.6 & 17.9 & 29.9 & 16.3 & 29.7 & 16.6 \\
Gemini 3.1 Pro & 96.3 & 96.0 & 53.5 & 36.1 & 26.7 & 18.1 & 35.7 & 31.3 & 30.7 & 17.6 \\
$^\dagger$Gemini 3.1 Pro & 97.2 & 97.0 & 100.0 & 100.0 & 100.0 & 100.0 & 100.0 & 100.0 & 100.0 & 100.0 \\
\midrule
\multicolumn{11}{c}{Conversational Voice Aesthetic Model (CVAM)} \\ \midrule
SFT$_{\textit{synth}}$ & \textbf{99.0} & \textbf{98.9} & 73.1 & 59.0 & 33.7 & \textbf{25.5} & 71.2 & 72.1 & 41.5 & 35.8 \\
$^\dagger$SFT$_{\textit{synth}}$ & 98.9 & 98.8 & 100.0 & 100.0 & 99.9 & 99.9 & 100.0 & 100.0 & 99.9 & 100.0 \\
SFT$_{\textit{synth}}$ + GRPO$_{\textit{human}}^{\textit{Frac}}$ & 98.8 & 98.7 & \textbf{77.4} & \textbf{64.0} & \textbf{35.0} & 20.5 & \textbf{73.7} & \textbf{73.4} & \textbf{45.0} & \textbf{41.3} \\
$^\dagger$SFT$_{\textit{synth}}$ + GRPO$_{\textit{human}}^{\textit{Frac}}$ & 98.8 & 98.7 & 99.9 & 100.0 & 99.9 & 99.9 & 100.0 & 100.0 & 99.9 & 99.9 \\
SFT$_{\textit{synth}}$ + GRPO$_{\textit{human}}^{\textit{KL}}$ & 98.8 & 98.7 & 73.7 & 59.6 & 33.0 & 19.1 & 71.9 & 72.7 & 42.6 & 36.2 \\
$^\dagger$SFT$_{\textit{synth}}$ + GRPO$_{\textit{human}}^{\textit{KL}}$ & 98.4 & 98.3 & 100.0 & 100.0 & 100.0 & 100.0 & 100.0 & 100.0 & 100.0 & 100.0 \\
\bottomrule
\end{tabular}}
\caption{Agreement with speaker self-reported gender and tool-measured objective fields. $^\dagger$\textit{Tools measure} provides the pitch and pacing labels and the evidence log, so those rows reach $100\%$ by construction. Only compare models without tool measure.}
\label{tab:objective}
\vspace{-0.5cm}
\end{table}

\section{Results} \label{sec:results}

We evaluate agreement with human majority-voted perceptual labels using accuracy and macro-F1 in Table~\ref{tab:perceptual}. Accuracy measures overall agreement, whereas macro-F1 weights all classes equally, including rare ones. All CVAM checkpoints outperform the evaluated open-source speech LLMs in average macro-F1. Comparisons among training methods further highlight the complementary roles of synthetic data warmup and human data alignment. Starting directly from the base model, GRPO on human labels (1.5k) and SFT on Gemini description + human labels reach only 34.7\% and 38.1\% F1, respectively. (For the latter, Gemini supplies the missing descriptions conditioned on the majority-voted perceptual labels.) In comparison, SFT on synthetic data (30k) alone achieves 43.5\% F1, providing a much stronger starting point for human alignment. Subsequent SFT on human labels raises F1 to 44.9\%, while GRPO with forward-KL rewards reaches 50.1\%. Thus, synthetic supervision provides an effective warmup, and human supervision delivers further gains, with GRPO outperforming continued SFT on humans.

Table~\ref{tab:objective} reports objective label prediction for completeness, as providing tool measures essentially solves this task. Nonetheless, the tool measures slightly improve perceptual label prediction as well (Table~\ref{tab:perceptual} with or without $\dagger$). If tools are prohibited (inference from raw audio), SFT followed by GRPO remains the most effective combination, achieving the best performance on most objective metrics.

\definecolor{optedcolor}{HTML}{E6F3FB}
\begin{table}[!h]
\centering
\small
\setlength{\tabcolsep}{3pt}
\resizebox{\columnwidth}{!}{%
\begin{tabular}{l | cc cc cc cc | cc}
\toprule
\multirow{2}{*}{GRPO$_\textit{human}$ Reward} & \multicolumn{2}{c}{Emotion} & \multicolumn{2}{c}{Arousal} & \multicolumn{2}{c}{Warmth} & \multicolumn{2}{c}{Confidence} & \multicolumn{2}{c}{Avg. Perceptual} \\
\cmidrule(lr){2-3} \cmidrule(lr){4-5} \cmidrule(lr){6-7} \cmidrule(lr){8-9} \cmidrule(lr){10-11}
 & Acc & F1 & Acc & F1 & Acc & F1 & Acc & F1 & Acc & F1 \\
\midrule
\rowcolor{gray!15} $^\dagger$SFT$_\textit{synth}$ \hspace{2pt} Baseline & 72.9 & 36.1 & 55.0 & 46.0 & 60.1 & 47.1 & 47.5 & 44.8 & 58.9 & 43.5 \\
\midrule
Acc on single-human labels & 84.7 & 35.2 & 77.7 & 53.9 & 74.7 & 51.7 & 60.3 & \textbf{52.9} & 74.4 & 48.4 \\
Acc on majority-voted labels & 86.6 & 38.5 & 77.6 & 48.5 & 78.2 & 51.7 & 64.4 & 49.5 & 76.7 & 47.0 \\
Fraction of shared labels & \textbf{88.1} & 33.0 & \textbf{78.8} & 52.5 & \textbf{78.7} & 53.6 & \textbf{65.7} & 51.1 & \textbf{77.8} & 47.6 \\
TVD on votes' distribution & 77.2 & 37.4 & 74.1 & 58.7 & 65.3 & \textbf{53.7} & 52.5 & 49.1 & 67.3 & 49.7 \\
KL on votes' distribution & 83.9 & \textbf{40.5} & 75.3 & \textbf{59.3} & 66.0 & 51.9 & 53.1 & 48.6 & 69.6 & \textbf{50.1} \\
Rev KL on votes' distribution & 83.9 & 35.8 & 75.8 & 55.6 & 64.9 & 44.2 & 49.9 & 43.0 & 68.6 & 44.7 \\
\bottomrule
\end{tabular}}
\caption{Study of GRPO rewards from the same SFT checkpoint. The first three are accuracy-based, and the last are distribution-based.}
\label{tab:reward}
\end{table}

To study the effect of reward design, Table~\ref{tab:reward} compares models trained with different GRPO rewards on human votes, all initialized from the same SFT checkpoint. Among them, vote-fraction rewards achieve the highest accuracy (77.8\%), whereas forward KL rewards achieve the highest macro-F1 (50.1\%). This trade-off supports our intuition that distribution-based rewards encourage coverage of less frequent human judgments, while vote-fraction rewards favor agreement with majority labels. TVD achieves similar F1 to forward KL, whereas reverse KL performs worse. Both vote-fraction and forward-KL models exceed Gemini 3.1 Pro and single-human-vs.-rest agreement in average accuracy and F1.

\begin{table}[!h]
\centering
\small
\setlength{\tabcolsep}{3pt}
\resizebox{\columnwidth}{!}{%
\begin{tabular}{l cc cc cc cc | cc}
\toprule
\multirow{2}{*}{Version} & \multicolumn{2}{c}{Emotion} & \multicolumn{2}{c}{Arousal} & \multicolumn{2}{c}{Warmth} & \multicolumn{2}{c}{Confidence} & \multicolumn{2}{c}{Avg. Perceptual} \\
\cmidrule(lr){2-3} \cmidrule(lr){4-5} \cmidrule(lr){6-7} \cmidrule(lr){8-9} \cmidrule(lr){10-11}
 & Acc & F1 & Acc & F1 & Acc & F1 & Acc & F1 & Acc & F1 \\
\midrule
Real & 82.0 & 34.0 & 76.0 & 52.3 & \textbf{68.6} & 45.9 & 51.0 & 45.0 & 69.4 & 44.3 \\
Voice-Cloned & \textbf{90.8} & \textbf{68.7} & \textbf{81.2} & 50.4 & 65.2 & 40.3 & 52.8 & 39.7 & \textbf{72.5} & 49.8 \\
Voice-Designed & 79.0 & 33.6 & 68.6 & \textbf{62.8} & 64.2 & \textbf{51.4} & \textbf{55.6} & \textbf{52.8} & 66.9 & \textbf{50.1} \\
\bottomrule
\end{tabular}}
\caption{Performance of CVAM ($^\dagger$SFT$_{\textit{synth}}$ + GRPO$_{\textit{human}}^{\textit{KL}}$) on real and synthetic speeches with the same (real human speech) context. Each version contributes 0.5k of the 1.5k test turns.}
\label{tab:byversion}
\end{table}

Table~\ref{tab:byversion} breaks down CVAM's performance on real, voice-cloned, and voice-designed responses (same text, different styles). Voice-designed speech achieves the highest overall F1, consistent with its more balanced labels. Voice-cloned speech achieves the highest emotion F1, possibly reflecting less ambiguous emotional expression; its predominance of neutral responses also favors accuracy. Real speech remains the most challenging overall, reflecting the complexity of natural, spontaneous expression. These differences show that performance varies with both the response version and the perceptual attribute being evaluated.

\begin{figure}[!h]
    \centering
    \includegraphics[width=\linewidth]{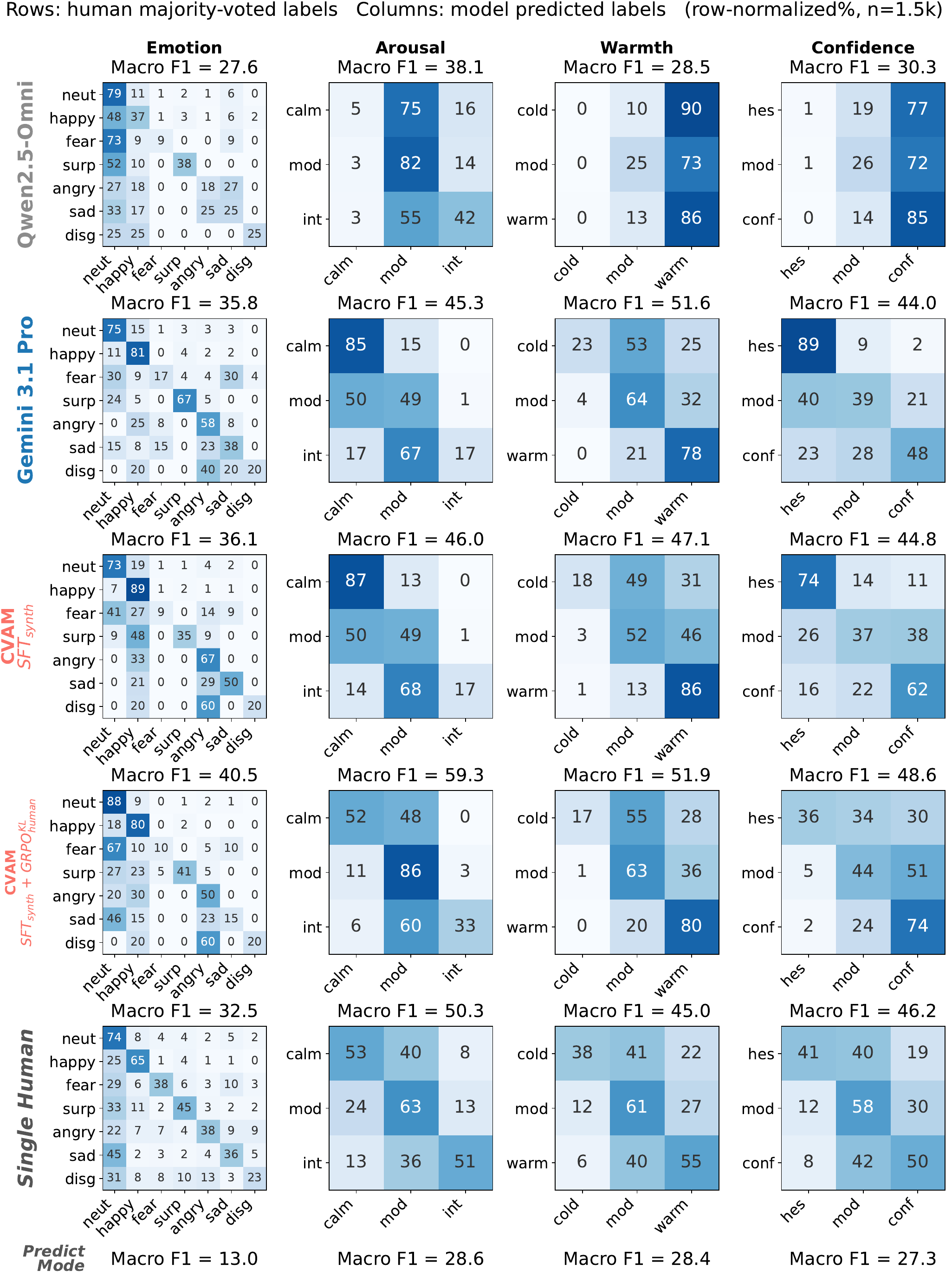}
    \caption{Confusion matrices and macro-F1 (above) for the perceptual labels on 1.5k test responses. Both SFT and GRPO stages improve macro-F1 over the base Qwen2.5-Omni. \textit{Predict Mode} is included as a baseline, which always outputs the most frequent label.}
    \label{fig:confusion}
\end{figure}

The confusion matrices in Fig.~\ref{fig:confusion} complement these aggregate results by visualizing predictions against human majority labels, alongside the base model, Gemini 3.1 Pro, and individual listeners. Synthetic SFT and subsequent GRPO each improve macro-F1 across all four perceptual labels. The matrices also reveal that these aggregate gains do not imply uniform improvements for every class, with some rare emotions remaining difficult to identify.

\noindent\rule{\linewidth}{0.4pt}

Together, these results support our two-stage approach to conversational voice aesthetic modeling: SFT on synthetic data followed by GRPO on human votes. GRPO reward design shapes the balance between majority-label accuracy and broader class coverage, with forward-KL rewards achieving the highest macro-F1. Spontaneous speech and rare perceptual classes remain challenges.

\section{Conclusion}
We introduce CVAM to describe the voice aesthetics of real and synthetic speech in a conversational context. With SFT on synthetic annotations and GRPO on human votes, CVAM outperforms evaluated speech LLMs and individual humans in agreement with human listeners. We explore rewards on perceptual labels against human votes: vote-fraction rewards favor majority-label accuracy, while distribution-based rewards improve macro-F1 and class coverage. Future directions include scaling human annotation to more samples and more votes per sample, and using CVAM as a reward model for expressive speech generation and conversational voice agents.

\newpage

\bibliographystyle{IEEEbib}
\bibliography{strings, refs_shorten}

\end{document}